\documentclass[mathleft
]{an}
\usepackage{graphicx}
\usepackage{times}
\usepackage{subfigure}
\usepackage{hyperref}
\usepackage{natbib}
\bibpunct{(}{)}{;}{a}{}{,}
\newcommand{\nat}{Nat}
\newcommand{\aap}{A\&A}                   

\newcommand{\cm}{\rm\thinspace cm}

\newcommand{\s}{\rm\thinspace s}
\newcommand{\ks}{\rm\thinspace ks}

\newcommand{\Ms}{\rm\thinspace Ms}

\newcommand{\keV}{\rm\thinspace keV}

\newcommand{\erg}{\rm\thinspace erg}

\newcommand{\ergcmps}{\hbox{$\erg\cm\ps\,$}}

\newcommand{\ps}{\hbox{$\s^{-1}\,$}}

\newcommand{\rg}{\rm\thinspace $r_\mathrm{g}$}

\begin{document}

\Pagespan{1}{}
\Yearpublication{2015}%
\Yearsubmission{2015}%
\Month{09}%
\Volume{}%
\Issue{}%

\title{Driving Extreme Variability --- \\
        Measuring the evolving coron\ae\ and evidence for jet launching in AGN}

\author{D.R. Wilkins\thanks{Corresponding author:
  \email{drw@ap.smu.ca}\newline}
}
\titlerunning{Measuring the evolving coron\ae\ in AGN}
\authorrunning{D.R. Wilkins}
\institute{
Department of Astronomy and Physics, Saint Mary's University, Halifax, NS. B3H 3C3. Canada}

\received{1 September 2015}


\abstract{%
Relativistically blurred reflection from the accretion disc provides a powerful probe of the extreme environments close to supermassive black holes; the inner regions of the accretion flow and the corona that produces the intense X-ray continuum. Techniques by which the geometry and extent of the corona can be measured through the observed X-ray spectrum are reviewed along with the evolution in the structure of the corona that is seen to accompany variations in the X-ray luminosity both on long and short timescales. Detailed analyses of the narrow line Seyfert 1 galaxies Markarian 335 and 1H\,0707$-$495, over observations with \textit{XMM-Newton} as well as \textit{Suzaku} and \textit{NuSTAR} spanning nearly a decade reveal that increases in the X-ray luminosity coincide with an expansion of the corona to cover a larger area of the inner accretion disc. Underlying this long timescale variability lie more complex patterns of behaviour on short timescales. Flares in the X-ray emission during a low flux state of Mrk~335 observed in 2013 and 2014 are found to mark a reconfiguration of the corona while there is evidence that the flares were caused by a vertical collimation and ejection of coronal material, reminiscent of an aborted jet-launching event. Measurements of the corona and reflecting accretion disc are combined to infer the conditions on the inner disc that lead to the flaring event.}

\maketitle

\section{Introduction}
Active galactic nuclei (AGN) are some of the most luminous objects seen in the Universe. Powered by the accretion of matter onto a supermassive black hole, they are extremely luminous sources of X-ray emission that is thought to originate from a `corona' of energetic particles surrounding the black hole and inner regions of the accretion flow.

The mechanism by which the corona is produced and by which energy is injected into it remains a mystery. It is possible that particles are accelerated by magnetic fields anchored to the ionised accretion disc undergoing reconnection events \citep[\textit{e.g.}][]{galeev+79,haardt+91,merloni_fabian} and that the intense X-ray continuum is produced by the inverse-Compton scattering of predominantly ultraviolet thermal seed photons emitted from the accretion disc up to X-ray energies \citep{sunyaev_trumper}. Understanding the characteristics and formation of the corona is vital to the understanding of how these incredibly luminous objects are powered.

The accretion flow is illuminated by the X-rays emitted from the corona \citep{george_fabian} and a characteristic reflection spectrum is produced \citep{ross_fabian}, shifted in energy and blurred by Doppler shifts (due to the orbital motion of the reflecting material) and gravitational redshift, shifting X-rays to lower energy the closer to the black hole they were emitted \citep{fabian+89,laor-91}. Detailed analysis of the reflected X-rays enables measurements to be made of the extreme environment around the black hole; the inner regions of the accretion flow and the corona. The reflected X-rays are most easily measured in Seyfert galaxies with little absorption along the line of sight to the nucleus (the `bare Seyferts'). 

The X-ray emission from AGN displays extreme variability \citep{leighly-99_1,turner+99}, in particular that from the narrow line Seyfert 1 (NLS1) galaxies \citep{ponti+2012}. The observed X-ray count rate will often vary by factors of two or three on timescales of just hours and sources will be seen to transition between states of high and low flux. If changes in the properties of the corona can be measured between epochs of high and low flux, it will give important insight into the process by which the corona is produced and energised by the accretion process.

In this article, the techniques by which the geometry and extent of the corona can be measured through analysis of the relativistically blurred reflection from the accretion disc are reviewed along with the evolution of coron\ae\ through periods of high and low flux that has been observed; both the long-timescale transitions from low to high flux states and short-lived flares seen during low flux states.

\section{Measuring the geometry of the corona}
\begin{figure*}
\centering
\subfigure[Emissivity, point source] {
\includegraphics[height=42mm]{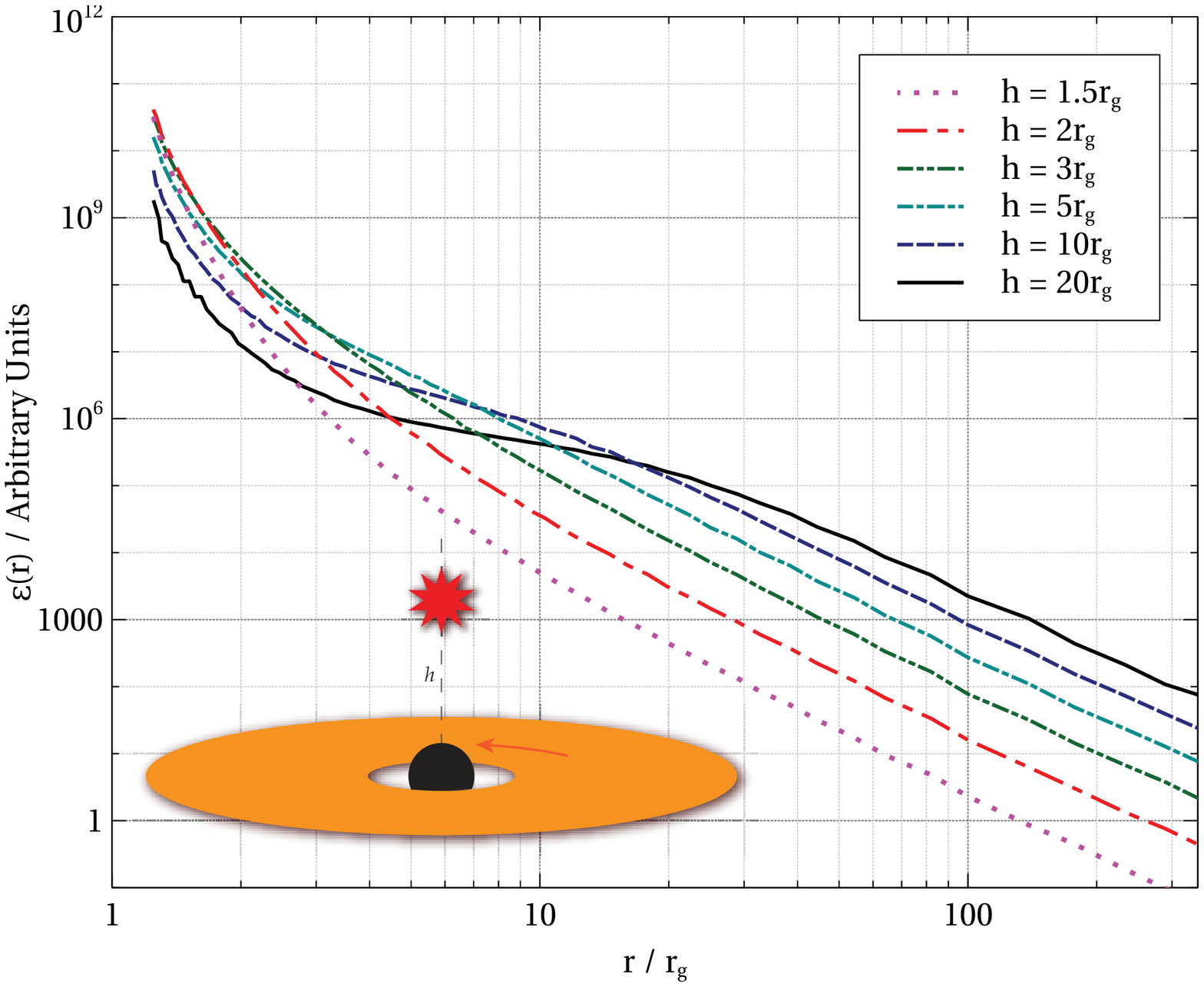}
\label{emis_theory.fig:point}
}
\subfigure[Emissivity, radially extended coron\ae] {
\includegraphics[height=42mm]{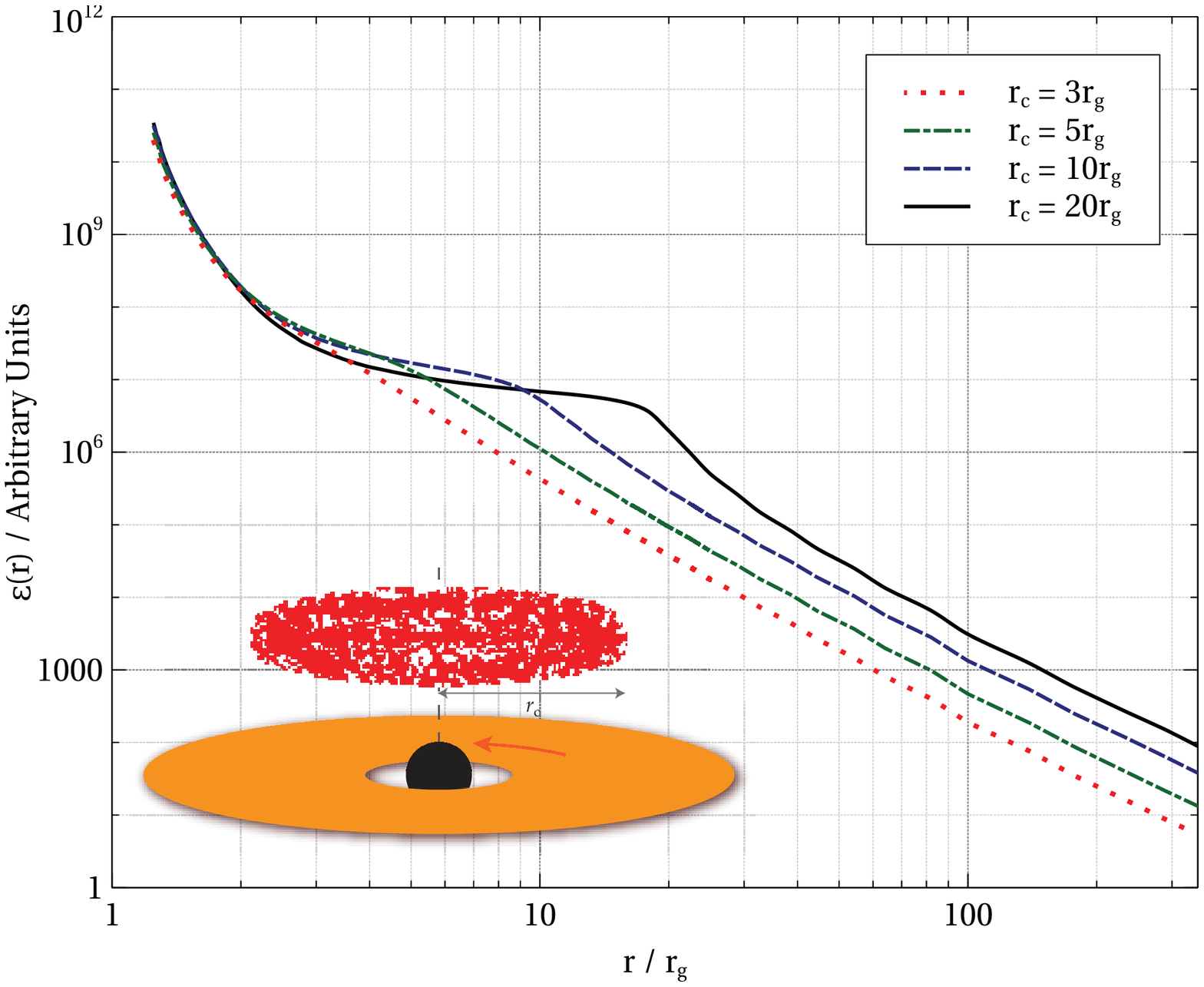}
\label{emis_theory.fig:ext}
}
\subfigure[Reflected and continuum fluxes, point source] {
\includegraphics[height=42mm]{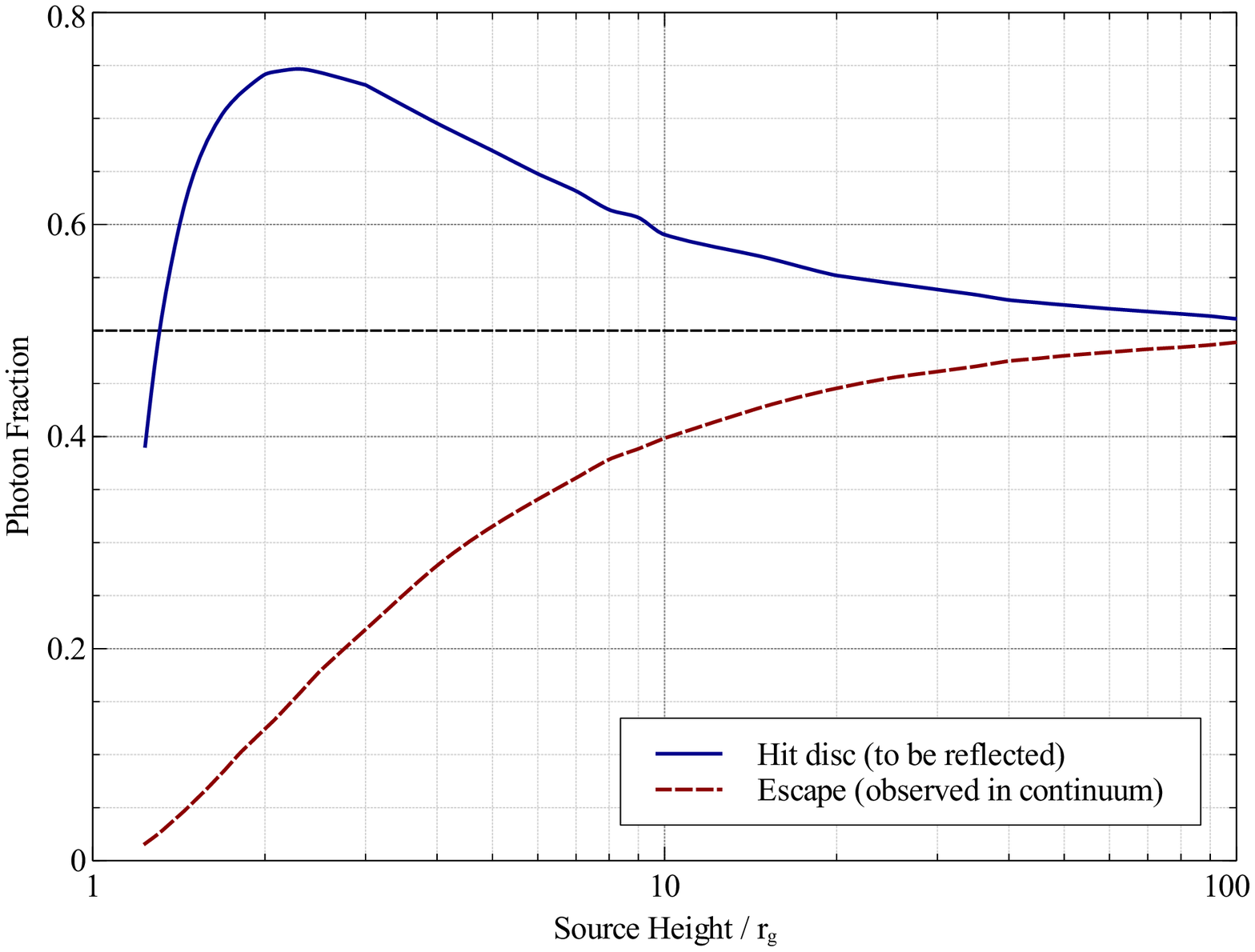}
\label{emis_theory.fig:reffrac}
}
\caption[]{\subref{emis_theory.fig:point} and \subref{emis_theory.fig:ext} Theoretical emissivity profiles for accretion discs illuminated by point sources and coron\ae\ extended radially over the surface of the disc.  \subref{emis_theory.fig:reffrac} The fraction of photons emitted from a point source at varying height that hit the disc (to be reflected) and escape (to be observed as part of the continuum), illustrating the rising reflection fraction for X-ray sources confined closer to the black hole.}
\label{emis_theory.fig}
\end{figure*}

\subsection{The accretion disc emissivity profile}
\label{emissivity.sec}
Where X-rays emitted from the corona illuminate the underlying accretion disc, the pattern of illumination they cast over the disc (the \textit{emissivity profile} of the accretion disc) reveals the location and geometry of the corona \citep{understanding_emis_paper}.

The accretion disc emissivity profile can be measured directly, exploiting the variation in Doppler shift (due to the orbital velocity) and gravitational redshift as a function of radius from the black hole. The reflection spectrum is decomposed into the contributions from successive radii in the disc and the contribution of each radius (hence the variation in illumination of the disc; the emissivity profile) is found through maximum likelihood fitting to the observed spectrum in the energy band encompassing the broadened iron K$\alpha$ emission line. Through this technique, \citet{1h0707_emis_paper} find that the emissivity profile of the disc in the NLS1 galaxy 1H\,0707$-$495 approximately takes the form of a twice broken power law, falling off steeply with index $>7$ over the inner regions of the disc, then flattening to almost a constant between $5\sim 35$\rg\ before falling off slightly steeper than $r^{-3}$ over the outer part of the disc. 

\citet{understanding_emis_paper} present a systematic analysis of the expected emissivity profiles for accretion discs illuminated by a range of point-like and extended coron\ae, derived from general relativistic ray tracing simulations. Profiles are shown in Fig.~\ref{emis_theory.fig}. In the case of an isotropic point source above the black hole, the emissivity profile approximately takes the form of a twice-broken power law \citep[see also][]{miniutti+03,dauser+13}. The emissivity falls off steeply over the innermost part of the disc, as $r^{-7}$ or steeper within 4\rg\ of the black hole. Rays are focused towards the black hole and hence onto the inner disc, while the coronal X-ray emission is strongly blueshifted into the frame of the material on the innermost orbits due to time dilation close to the black hole (in commonly-adopted models of relativistically broadened emission lines, the emissivity profile is defined in the frame of the emitting material). The profile then flattens until $r \sim h$, whereafter the profile falls off slightly steeper than $r^{-3}$ over the outer disc.

In order to reproduce both the extreme steepening observed over the inner disc and the break from the flattened profile to $r^{-3}$ at a few tens of gravitational radii from the black hole, the corona must extend radially over the inner part of the disc at a low height. In this case, the profile once again takes the form of a twice broken power law, flattened over the middle section and then breaking to $r^{-3}$ over the outer disc at a radius corresponding to the maximum radial extent of the source. In this case, the emission of X-rays at a low height over the central regions produces the extreme steepening over the inner part of the accretion disc in addition to the extended flattened portion of the profile. The emissivity profile of the accretion disc in 1H\,0707$-$495 is best explained by illumination of the disc by a corona extended radially over the disc to around 35\rg, but at a low height from the surface of the disc. Similarly, the corona in the NLS1 galaxy IRAS\,13224$-$3809 was found to extend to around 10\rg\ over the disc \citet{iras_fix} .

\begin{figure*}
\centering
\includegraphics[width=170mm]{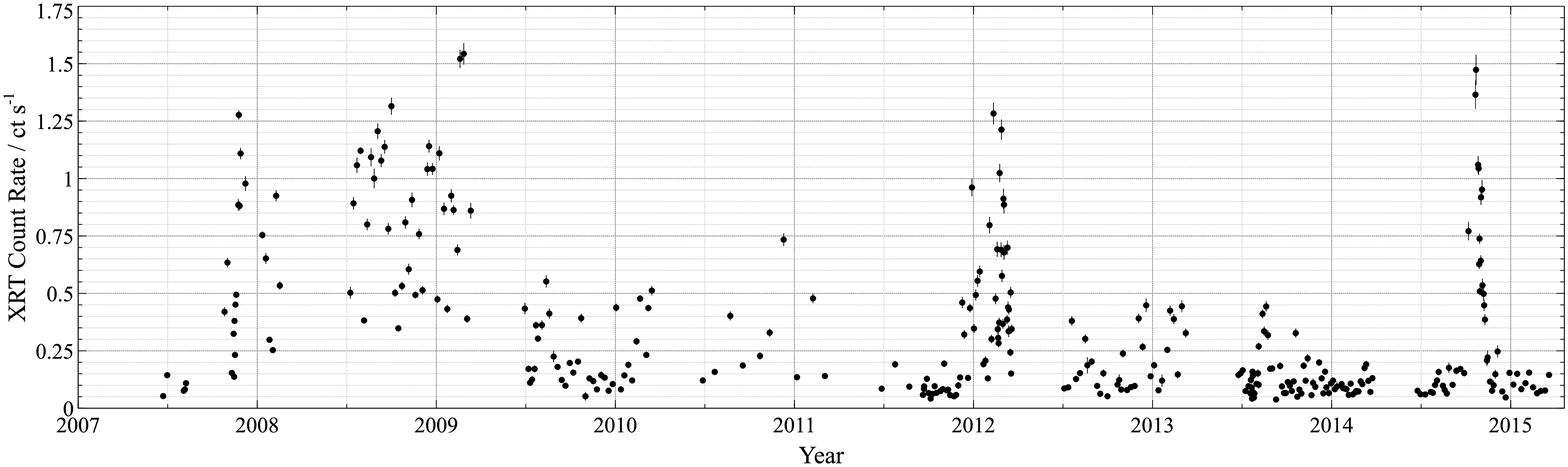}
\caption[]{The 0.5-5\keV\ X-ray light curve of the NS1 galaxy Mrk~335 measured with the \textit{Swift XRT}, illustrating the extreme variability exhibited by this source, including epochs of high and low flux as well as X-ray flares.}
\label{swift_lc.fig}
\end{figure*}

\subsection{The reflection fraction}
As the corona becomes confined to a more compact region around the black hole, a greater fraction of the emitted rays are focused towards the black hole and hence onto the inner regions of the accretion disc. Fewer rays are able to escape to be observed directly in the continuum, increasing the reflection fraction (the ratio of the total reflected to directly observed continuum flux) from unity, expected for the case of an isotropic point source illuminating an accretion disc in the absence of the black hole where 50 per cent of photons emitted from the source will travel down to the disc while 50 per cent escape to be observed directly. \citep{1h0707_jan11}. Measurements of the reflection fraction from the X-ray spectrum of AGN can therefore give a further indication of the extent of the corona and its proximity to the black hole.

The reflection fraction is not, na\"ievly, expected to drop below unity. Half of the emitted continuum photons will hit the disc while half will be able to escape to form the continuum in the absence of gravitational light bending which enhances reflection. \citet{beloborodov}, however, show that the upward beaming of emission from a corona moving at a relativistic velocity away from the disc can reduce the reflection fraction below unity. Such low reflection fractions suggest that the X-ray emitting corona is outflowing away from the disc at a mildly relativistic velocity.

\section{The evolving corona of Mrk 335}
Markarian 335 (Mrk~335) is a particularly fascinating example of a narrow line Seyfert 1 galaxy at redshift $z = 0.026$.  While early observations found a bright X-ray source in a high flux state, commencement of monitoring of this source with the \textit{Swift} satellite in 2007 revealed a factor of 10 drop in flux since the 2006 (and earlier) high flux observations \citep{grupe+07}. Mrk~335 has since been observed in intermediate \citep{grupe+12,gallo+13} and further low flux states \citep{gallo+14,parker_mrk335}. This extreme variability, illustrated in Fig.~\ref{swift_lc.fig}, makes Mrk~335 an ideal target to study the changes AGN coron\ae\ undergo that lead to the extreme variability in their X-ray emission.

X-ray spectra of Mrk~335 are well-described by the relativistically blurred reflection of the coronal X-ray emission from the accretion disc, from the early high-flux observations with \textit{XMM-Newton} \citep{crummy+06}, to the low \citep{grupe+08,parker_mrk335,gallo+14} and intermediate flux \citep{gallo+13} observations. While the spectrum is dominated by reflection from the accretion disc, the presence of a warm outflowing absorber has been detected during several of the epochs and has been attributed to a wind from disc \citep{longinotti+13}.

\subsection{Long timescale variability}
\begin{figure}
\centering
\includegraphics[width=80mm]{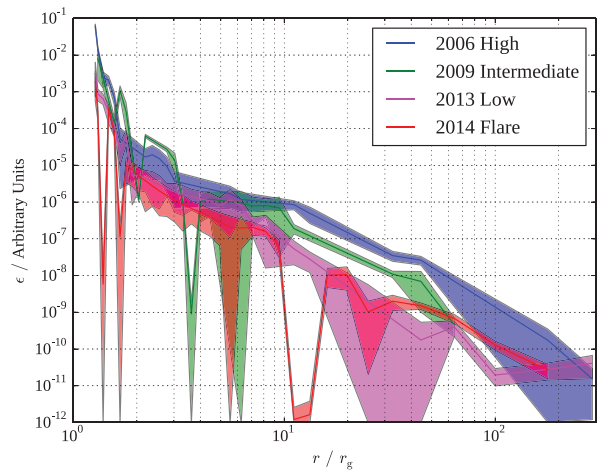}
\caption[]{The emissivity profile of the accretion disc in Mrk~335 during high, intermediate and low flux epochs as well as during a flare observed during the low state in 2014, measured by decomposing the relativistically blurred reflection from the disc into the contributions from successive radii and fitting the relative contributions to the observed spectra.}
\label{emis_epochs.fig}
\end{figure}

The emissivity profile of the accretion disc during epochs of high, intermediate and low flux taken with \textit{XMM-Newton} in 2006 and 2009 and \textit{Suzaku} in 2013, respectively was measured following the procedure of \citet{1h0707_emis_paper} in order to understand the underlying changes in the structure of the corona that gave rise to the extreme variability seen in the X-ray emission from this object \citep{mrk335_corona_paper}. The values of of the key model parameters from which the properties of the corona are inferred during each of the observations are shown in Table~\ref{par.tab} and the emissivity profile of the accretion disc measured during each epoch is shown in Fig.~\ref{emis_epochs.fig}.

\begin{table*}
\centering
\caption{\label{par.tab}The best-fitting values of key model parameters describing the X-ray continuum emission and relativistically blurred reflection from the accretion disc during epochs of high, intermediate and low flux of the NLS1 galaxy Mrk~335 as well as over the course of a flare during the 2013 low flux observation. In addition to the the reflection from the disc, the model includes absorption from outflowing ionised material and the unblurred reflection from distant material.}
\def\arraystretch{1.5}
\begin{tabular}{|ll|c|c|c|ccc|c|}
  	\hline
   	\textbf{Component} & \textbf{Parameter} & \textbf{2006 High} & \textbf{2009 Int.} & \multicolumn{4}{c|}{\textbf{2013 Low}} & \textbf{2014 Flare}\\
	& & & & All & Before & During & After Flare & \\
	\hline
	\multicolumn{2}{|l|}{$F_{0.5-10\mathrm{keV}}$ / $10^{-12}$\ergcmps} & $41$ & $7.9$ & $3.8$ & $3.6$ & $6.6$ & $3.5$ & $11$ \\
	\hline
	Continuum & $\Gamma$ & $2.523_{-0.010}^{+0.011}$ &  $1.90_{-0.02}^{+0.02}$ & $1.91_{-0.07}^{+0.04} $ & $1.77_{-0.01}^{+0.01}$ & $1.94_{-0.04}^{+0.04}$ & $1.81_{-0.01}^{+0.01}$ & $2.43_{-0.09}^{+0.05}$ \\
	\hline
	Disc reflection & $r_\mathrm{in}$ / \rg & $1.235^{+0.003}$  & $1.24_{-0.05}^{+0.14}$ & $1.25_{-0.02}^{+0.03}$ & $1.26_{-0.02}^{+0.02}$ & $1.25_{-0.01}^{+0.01}$ & $1.81_{-0.01}^{+0.01}$ & $1.235^{+0.15}$ \\
	& $\xi$\,/\,\ergcmps & $58_{-4}^{+12}$  & $250_{-20}^{+30}$ & $13_{-5}^{+7}$ & $25_{-7}^{+27}$ & $<1.1$ & $9_{-6}^{+12}$ & $<3.2$\\
	& $R_{\,0.1-100\,\mathrm{keV}}$ & $1.3_{-0.2}^{+0.5}$ & $1.8_{-0.3}^{+0.4}$ & $6_{-3}^{+4}$ & $9_{-6}^{+7}$ & $1.8_{-1.5}^{+1.2}$ & $14_{-12}^{+18}$ & $0.41_{-0.15}^{+0.15}$ \\
	\hline
	Goodness of fit & $\chi^2 / \nu$  & 1.04 & 1.00 & 1.04 & 1.06 & 1.01 & 1.06 & 1.04 \\
	\hline
\end{tabular}
\end{table*}

During each epoch, the emissivity profile is found to take the approximate form of a twice-broken power law falling off steeply over the inner part of the accretion disc, with the power law index around 9 in each case, while tending to a constant power law slope, slightly steeper than $r^{-3}$ over the outer parts of the disc following a flattened portion of the profile over the middle part of the disc. The observed forms of the emissivity profile suggest that during each of these epochs, the X-ray continuum emission is arising from a corona extended radially at a low height over the surface of the disc.

Variation in the extent of the flattened section of the emissivity profile suggests a variation in the radial extent of the corona. The emissivity profiles suggest that the corona is more extended during the high flux epochs than when the flux is lower. Fitting the observed profile of iron K$\alpha$ line with a continuous, twice-broken power law emissivity profile finds the best fitting outer break radius (\textit{i.e.} the radial extent of the corona over the surface of the disc) contracted from $26_{-7}^{+10}$\rg\ during the high flux epoch to $<12$\rg\ in the intermediate flux state and $<5$\rg\ during the low flux observation (at the 90 per cent confidence level). The contraction of the corona from the high to low flux epochs is also seen in the variation in the reflection fraction. As the corona contracts from the extended structure over the surface of the disc in 2006 to the confined region around the black hole in 2013, a greater fraction of the X-rays emitted are focused towards the black hole and hence onto the inner regions of the accretion disc, rather than being able to escape to be observed as part of the continuum.

\subsection{Coronal variability across AGN}
Evidence has previously been found for the expansion of the X-ray emitting corona as the X-ray count rate increases. Between early observations in 2002 and 2011, the X-ray flux received from the NLS1 galaxy 1H\,0707$-$495 remained at a fairly consistent mean level, although frequently varying around this by factors of two or three over timescales of just hours. \citet{1h0707_var_paper} divided up more than 1.3\Ms\ of observations of 1H\,0707$-$495 made with \textit{XMM-Newton} into periods of high and low flux.

When the X-ray luminosity was greater, the corona was, on average, found to extend over a larger portion of the inner disc. The outer break radius in the emissivity profile from the flattened middle section to the inverse-cube fall-off over the outer disc was found to increase from $23_{-3}^{+3}$\rg\ during the lowest flux intervals to $30_{-3}^{+5}$ during the highest. The variation in the corona between the high, intermediate and low flux epochs of Mrk~335 would appear to be a detection of greater significance of the same process occurring on a greater scale, manifesting a much greater change in X-ray flux in the form of a `state change' rather than the much more frequent fluctuation in flux seen in 1H\,0707-495. 

The 2013 low flux epoch of Mrk~335 is in many ways similar to the sudden drop in X-ray flux seen from 1H\,0707-$495$ in 2011. The X-ray flux from 1H\,0707-$495$ was seen to drop by more than an order of magnitude. The X-ray emission was dominated by relativistically blurred reflection from the accretion disc with little or no contribution from directly observed continuum emission and a steeply falling emissivity profile. The steep emissivity profile and extremely high reflection fraction are suggestive of the corona having collapsed into a very confined region, within just 2\rg\ of the black hole \citep{1h0707_jan11}.

\citet{kara_iras_lags} find evidence for the expansion of the corona through the reverberation time lag measured during a high flux period of 2011 X-ray observations of the NLS1 galaxy IRAS\,13224$-$3809. It was found that as the X-ray count rate increased by a factor of $4\sim 5$, the time lag between continuum-dominated and reflection-dominated energy bands had increased by a factor of around 3. X-ray reverberation time lags are sensitive to the vertical location or extent of the corona above the plane of the accretion disc while being relatively insensitive to the radial extent of the corona \citep{lag_spectra_paper} as opposed to the accretion disc emissivity profiles measured in 1H\,0707$-$495 and Mrk~335, hence there is evidence for the vertical expansion of the corona as well as radial expansion over the surface of the disc as AGN transition from epochs of low to high flux.

\subsection{Flaring and reconfiguration of the corona}
During the 2013 low flux state observation with \textit{Suzaku}, a flare was observed in the X-ray emission from Mrk~335 during which the count rate doubled for approximately 100\ks\ before returning to its pre-flare level. \citet{mrk335_corona_paper} conduct a detailed analysis of the X-ray spectrum and the accretion disc emissivity profile during time periods before, during and after the flare while \citet{gallo+14} present a time-resolved analysis of the reflection spectrum arising from the accretion disc. The variation in key spectral parameters from which the evolution of the corona and the cause of this flare can be studied in time periods before, during and after the flare are shown in Table~\ref{par.tab}.

\begin{figure}
\centering
\includegraphics[width=80mm]{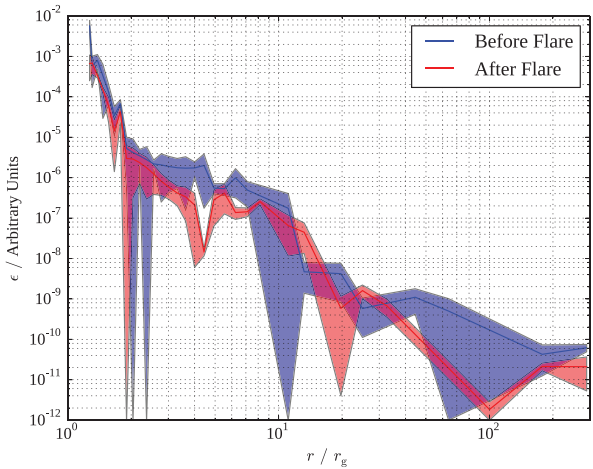}
\caption[]{Comparison of the accretion disc emissivity profile before and after a flare during the 2013 low flux observation of Mrk~335.}
\label{emis_flare.fig}
\end{figure}

Fig.~\ref{emis_flare.fig} compares the emissivity profile of the accretion disc in Mrk~335 before and after flare. Most notably, the emissivity profile before the flare shows flattening out to 5\rg\ which is not present after the flare. The flare marks a reconfiguration of the corona from being slightly extended over the surface of the accretion disc (but to no more than 5\rg, significantly less than in during the intermediate and high flux epochs) to being much more compact, confined within 2\rg\ of the black hole after the flare for no flattening to be seen in the emissivity profile.

\begin{figure*}
\centering
\includegraphics[width=160mm]{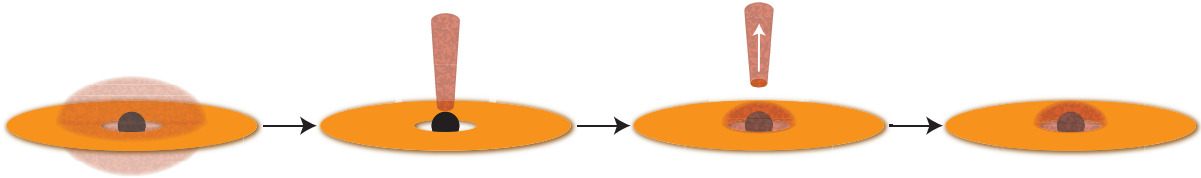}
\caption[]{The evolution of the corona, inferred from measurements of the accretion disc emissivity profile and reflection fraction, over the course of flares seen during the low flux state of Mrk~335.}
\label{flare_evolution.fig}
\end{figure*}

During the flare itself, fitting the emissivity profile of the accretion disc as a once-broken power law and obtaining the probability distribution of the outer slope in each case suggests that from the time period before the flare to that during and after, the emissivity profile is steepening \citep{mrk335_corona_paper}. The steeping emissivity profile and dramatic drop in the reflection fraction during the flare allow the flare to be interpreted as the collimation of the X-ray emitting corona into a vertically extended structure moving upward at a mildly relativistic velocity. This collimated `jet-like' state of the corona, however, cannot be sustained and the X-ray emitting corona collapses back down to a much more confined region around the black hole as the jet-launching is aborted.

Combining these measurements with the evolution of properties of the accretion disc found through time-resolved analysis of the X-ray reflection spectrum \citep{gallo+14}, it is possible to infer the conditions on the inner regions of the accretion flow that gave rise to the flare. In particular, as the flare declined, the ionisation parameter of the disc was found to increase (\textit{i.e.} the disc was becoming more ionised). This gave rise to an approximately linear correlation between the X-ray flux in the 0.7-0.9 and 7-10\keV\ bands while no correlation was seen before and during the rise of the flare. During this time, the total flux in the reflected component of the spectrum was found to be decreasing, suggesting that while the disc was becoming more ionised, it was receiving less ionising flux from the corona. In the absence of other processes ionising the disc, the ionisation of the disc was therefore being suppressed during the flare by an increase in the disc density.

Such a situation has been suggested in general relativistic magneto-hydrodynamic simulations of magnetically choked accretion flows \citep{mckinney+2012}. Large-scale poloidal magnetic fields are accreted inwards through the ionised disc plasma, building up on the inner regions of the disc. They cause the inner part of the disc to become compressed and form a barrier, choking the mass accretion rate (which may explain the low flux state). When a critical magnetic flux density has built up on the inner disc, the magnetic field undergoes an inversion ejecting material from the inner regions. This could be responsible for the ejection of the collimated corona during the flare while the field responsible for compressing the inner part of the disc is lost allowing the density to once again drop and the inner regions to re-ionise.

Monitoring of Mrk~335 with the \textit{Swift} satellite revealed a much greater flare in 2014, lasting around 10 days. At the peak, the 0.3-5\keV\ count rate measured with the \textit{Swift XRT} had increased from that in the low flux state by a factor of 10. A target of opportunity observation was triggered with \textit{NuSTAR}, catching the decline of the flare, although the count rate was still at 60 per cent of its peak level. Simultaneous analysis of the X-ray spectra measured by the \textit{Swift XRT} and \textit{NuSTAR} revealed that the reflection fraction had dropped to the extremely low level of $0.41_{-0.15}^{+0.15}$. Decomposing the reflection spectrum into the contributions from successive radii on the disc revealed a steeply falling emissivity profile as was found during the 2013 low flux epoch (shown in Fig.~\ref{emis_epochs.fig}), with no evidence for flattening over the middle part of the disc.

These observations can be reconciled if the 2014 flare was caused by the same mechanism as that seen during the 2013 low flux observation, but on a greater scale. The corona becomes vertically collimated and is ejected at a relativistic velocity at the peak of the flare, beaming the continuum emission away from the disc. This leads to the low reflection fraction that is observed. By the time the \textit{NuSTAR} observation was made on the decline of the flare, the base of the corona had collapsed down to a confined region around the black hole, as was found after the 2013 flare. The accretion disc was illuminated predominantly by this compact corona that gives rise to the steeply falling emissivity profile. The coronal material, however, was ejected at a sufficient velocity during the flare that continuum emission is still seen from the upper part of the outflowing structure. This emission is beamed upwards and does not contribute significantly to the illumination of the disc. The proposed scheme giving rise to the flares is shown in Fig.~\ref{flare_evolution.fig}.

\section{Conclusions}
Where relativistically blurred reflection is detected from accretion discs in AGN, it provides a powerful probe of the inner regions of the accretion flow and the geometry of the X-ray emitting corona. Measurement of the emissivity profile of the accretion discs in a number of Seyfert galaxies reveals the corona to be extended over the inner regions of the accretion disc to a few tens of gravitational radii while extending to a low height vertically (just a few gravitational radii) above the plane of the disc.

Changes in the accretion disc emissivity profile between high and low flux states seen in single AGN, specifically in the NLS1 galaxies Mrk~335 and 1H\,0707$-$495, suggest that the corona expands to a greater extent over the disc during periods of high flux while collapsing to a confined region around the black hole as the flux drops.

While the long timescale variability and the regular fluctuations in X-ray flux appear to be due to the expansion and contraction of the corona over the surface of the disc, transient flares seen during the present low flux state of Mrk~335 coincide with the vertical collimation and ejection of the corona, suggestive of aborted jet-launching events. Flares were found to mark the reconfiguration of the corona from a slightly extended structure to a compact, collapsed structure close to the black hole as the outflowing, collimated corona cannot be sustained.

Measurements of the changing geometry of the corona that produces the intense X-ray continuum AGN through high and low flux states allow observations to be compared to the predictions of detailed theoretical simulations of accretion flows, coron\ae\ and jet-launching. This provides important insight into how energy is released from black hole accretion flows to produce the intense X-ray emission and potentially the powerful jets that are observed.

\acknowledgements
DRW is supported by a CITA National Fellowship. Thanks must go to Dirk Grupe for the \textit{Swift} monitoring data and to Luigi Gallo, Kirsten Bonson and Andy Fabian for insightful discussions during this research.



\bibliographystyle{mnras3}
\bibliography{agn}

\end{document}